\documentclass[aps,showpacs,prl,twocolumn,groupedaddress]{revtex4}
\usepackage{graphicx}

\begin{document}

\title{Adhesion of membranes via actively switched receptors} 
\author{Bartosz R\'{o}\.{z}ycki$^{1,2}$, Reinhard Lipowsky$^{1}$ and Thomas R.\ Weikl$^{1}$}

\affiliation{$^{1}$\mbox{Max-Planck-Institut f\"{u}r Kolloid- und 
Grenzfl\"{a}chtenforschung, 14424 Potsdam, Germany} \\ 
$^{2}$\mbox{Instytut Fizyki Teoretycznej, Uniwersytet Warszawski,
Ho\.za 69, 00 681 Warszawa, Poland}}

\begin{abstract}
We consider a theoretical model for membranes with adhesive receptors, or stickers, that are actively switched between two conformational states. In their `on'-state, the stickers bind to ligands in an apposing membrane, whereas they do not interact with the ligands in their `off'-state. We show that the adhesiveness of the membranes depends sensitively on the rates of the conformational switching process. This dependence is reflected in a resonance at intermediate switching rates, which can lead to large membrane separations and unbinding. Our results may provide insights into novel mechanisms for the controlled adhesion of biological or biomimetic membranes.
\end{abstract}


\maketitle

{\em Introduction} --
Fluid lipid bilayers are the fundamental building blocks of biological membranes and, as such, have been studied extensively \cite{lipowsky95}. As a consequence, many equilibrium properties of lipid membranes are now well understood. However, biological membranes are often out of equilibrium, driven by embedded active components or by the coupling to the dynamic cytoskeleton.  These active processes affect the membrane behavior, e.g.~by enhancing membrane shape fluctuations \cite{prost96,manneville99,ramaswamy00} or lateral diffusion \cite{granek99}. The enhancement of shape fluctuations has been theoretically predicted for membranes with ion pumps as active force centers \cite{prost96}, and experimentally verified for giant lipid vesicles with embedded bacteriorhodopsin pumps \cite{manneville99}. Other active membrane proteins have been suggested to couple to the local thickness \cite{sabra98} or curvature \cite{chen04} of the membranes.

\begin{figure}[b]
\begin{center}
\resizebox{0.95\columnwidth}{!}{\includegraphics{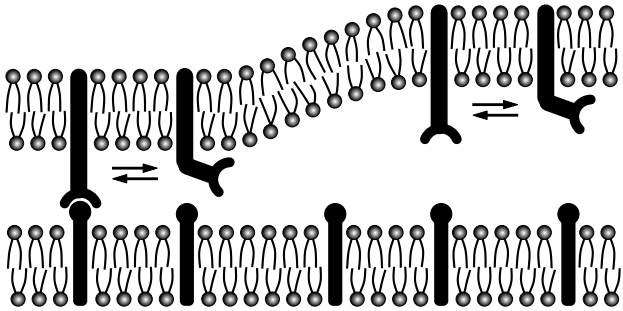}}
\caption{A membrane with active stickers (top) adhering to an apposing membrane. In one of their two conformations (stretched), the stickers are `on' and can bind to ligands in the second membrane. The stickers are actively switched between their conformations, with on- and off-rates that are independent of their binding state. Switching off a bound sticker (left) requires an energy input equivalent to the sticker binding energy, which keeps the system out of equilibrium.
\label{cartoon}}
\vspace*{-0.3cm}
\end{center} 
\end{figure}

In this letter, we consider another type of active membrane process. The active components embedded in these membranes are adhesion molecules, or `stickers', that can be switched between two states: an `on'-state in which the stickers interact with a second membrane, and an `off'-state  in which this interaction is negligible (see Fig.~\ref{cartoon}). An important biological example of actively switched adhesion molecules are integrins. In one of their conformations, the integrin molecules are extended and can bind to ligands present in an apposing membrane \cite{takagi02kim03}. In another conformation, the molecules are bent and, thus, deactivated. The transitions between these conformations are triggered by signaling cascades in the cells, which typically require energy input, e.g.~{\em via} ATP.  In biomimetic applications, conformational transitions of appropriately designed stickers may also be triggered by light \cite{moeller98abbott99ichimura00}.

In an equilibrium situation, the adhesiveness of a biomembrane only depends on the concentration and binding energy of the sticker molecules \cite{lipowsky96weikl01}. We show here that the  adhesiveness of membranes with active stickers also depends strongly on the switching rates of the stickers. 

\begin{figure}[b]
\begin{center}
\resizebox{0.85\columnwidth}{!}{\includegraphics{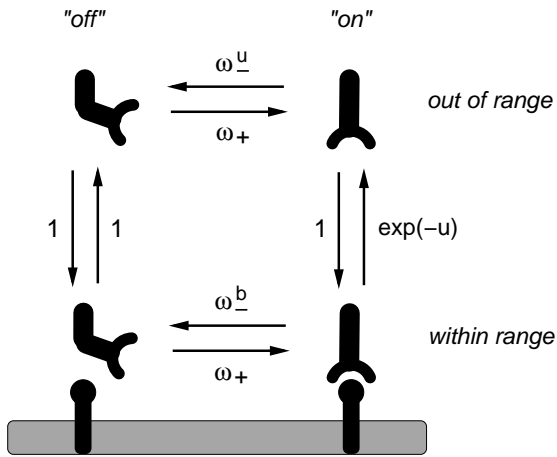}}
\caption{A simple four-state model of an active sticker. In this model, the sticker can attain only two different separations from its ligand, and two conformational states, `on' or `off'. At close separation, an `on'-sticker is bound to its ligand with binding energy $U=u k_B T$. The transition rates between the four states are given in units of $1/\tau$, where $\tau$ is a typical timescale for the local membrane relaxation between the two separations. The dimensionless rates $\omega_{-}^b$ and $\omega_{-}^u$ for switching a sticker off may in general depend on whether the sticker is bound or unbound, whereas the `on'-rate $\omega_{+}$ is taken to be independent of the separation.  
\label{fourstate}}
\vspace*{-0.3cm}
\end{center} 
\end{figure}

{\em Model} --
Before considering extended membranes with active stickers, let us first focus on a single sticker. 
In the simple example shown in Fig.~\ref{fourstate}, the sticker can only attain two different separations from its ligand. The sticker is within binding range of the ligand at close separation, and out of binding range at a larger separation. At both separations, the sticker can be in either of its two conformations, `on' or `off'. In this one-sticker model, we assume that the change in elastic energy of the membrane, which is associated with the change of separation, is negligible. The energy of an `off'-sticker is then independent of the separation, which implies that the rate for the transition from the larger to closer separation is identical to the transition rate for the reverse process. In the dimensionless units used in this example, the two transition rates are equal to 1. An `on'-sticker, in contrast, gains the dimensionless binding energy $u$ when bound to the ligand. If we adopt standard Metropolis dynamics, the dimensionless unbinding rate of this sticker is $\exp(-u)$, whereas the rate for the reverse process is 1. 

Let us now consider the transition rates between the `on'- and `off'-state. The rate $\omega_+$ for switching the sticker on is independent of the membrane separation if we assume that this transition is thermally activated and that the corresponding energy barrier is not affected by the presence of the ligand. However, the off-rates $\omega_{-}^{b}$ for a bound sticker and $\omega_{-}^{u}$ for an unbound sticker in general can be different. In equilibrium, the transition rates between the states have to obey detailed balance, which implies that the total rate $\omega_{-}^{b} \omega_{+}$ for passing through the four states in a clockwise direction has to be equal to the total rate $\exp(-u)\omega_{-}^{u} \omega_{+}$ for the counter-clockwise cycle. This leads to the equilibrium condition $\omega_{-}^{b}= \exp(-u)\omega_{-}^{u}$. 

In the following, we will consider a situation in which the stickers are switched off by an active process that violates detailed balance and, thus, the equilibrium condition above. The active process requires energy input, e.g.~via ATP hydrolysis. The energy that is set free by the hydrolysis of an ATP molecule typically is significantly larger than the binding energy of an adhesion molecule. Therefore, it seems not unreasonable to assume that the rate of an ATP-driven off-switching process is independent of whether the sticker is bound or not. We then have  $\omega_{-}^{b}=\omega_{-}^{u}\equiv \omega_{-}$.\cite{footnote} At small ATP concentration, the `off'-rate $\omega_{-}$ should be proportional to this concentration. 

Let us now consider a simple, discrete model for a membrane with active stickers that interact with ligands in an apposing membrane. In this model, a `reference plane' parallel to the membranes is discretized into a square lattice with lattice spacing $a$. At each site $i$, a sticker molecule is present that can be in the state `on' ($n_i=1$) or `off' ($n_i=0$). The sticker distribution within the membrane thus is approximated by a square-lattice distribution. The conformations of the membranes are described by the local membrane separation $l_i\ge0$ at the sticker sites $i$. The membrane separation is nonnegative since the membranes cannot penetrate each other. 

The configurational energy of the membranes at time $t$ has the form
\begin{equation}
H\left\{l, n(t)\right\} = \sum_{i} \left[\frac{\kappa}{2 a^2} \left(\Delta_{d} l_{i}\right)^{2} + n_{i}(t) V(l_{i}) \right]
\label{hamiltonian}
\end{equation}
The first term in this configurational energy represents the bending energy of the membranes. Here, $ \kappa = \kappa_{1} \kappa_{2} / (\kappa_{1} + \kappa_{2})$ is the effective bending rigidity of the membranes with rigidities $\kappa_1$ and $\kappa_2$, and the discretized Laplace operator $\Delta_{d} l_{i} = \Delta_{d} l_{x,y} = l_{x+1,y} + l_{x-1,y} + l_{x,y+1} + l_{x,y-1} - 4l_{x,y}$ captures the local mean curvature of the separation field $l$.  The second term is the time-dependent interaction energy of the membranes. If the sticker at site $i$ is `on' at time $t$, then $n_i(t)$ is 1 and the membranes interact with the short-ranged sticker potential $V(l_i)$. For simplicity, we model the short-ranged sticker interaction by the square-well potential $V(l_i) = -U$ for $0\le l_i \le l_r$ and  $V(l_i) = 0$ for $l_i>l_r$ with characteristic binding energy $U>0$ and potential range $l_r$. To reduce the number of parameters, we use the rescaled separation field $z_i = (l_i/a)\sqrt{\kappa/(k_B T)}$ in our simulations. In the following, the rescaled potential range $z_r = (l_r/a) \sqrt{\kappa/(k_B T)}$ of the stickers has the fixed 
value $z_r = 0.1$. 

The dynamics of the model here is simulated via a master equation with discrete time steps $\tau$. In a time step $\tau$, the rescaled separation $z_i$ at each lattice site $i$ is shifted to a value $z_i'$ with standard Metropolis transition probabilities: the `new' value $z_i'$ is adopted with probability 1 if the `new' conformation is lower in energy, and adopted with probability $\exp[-(H'-H)/(k_B T)]$ if the energy $H'$ of the `new' conformation is higher than the energy $H$ of the `old' conformation. Below, we choose $z_i'=z_i+ \zeta\rho$ where $\rho$ is a random number between -1 and 1, and $\zeta=0.5$ is the step size. 

To ensure consistency with versions of the model that are continuous in time \cite{preprint}, the sticker interaction energy is averaged over each time step $\tau$. If the sticker was always `on' in the interval $\tau$, the interaction energy is $V(z_i)$. If the sticker was always `off', the interaction energy is 0. If it was `on' during a fraction $q$ of the interval $\tau$, the interaction energy is $q V(z_i)$. The membrane separation field thus evolves in a stochastic sticker potential that is averaged over the time steps $\tau$.

\begin{figure}
\begin{center}
\resizebox{\columnwidth}{!}{\includegraphics{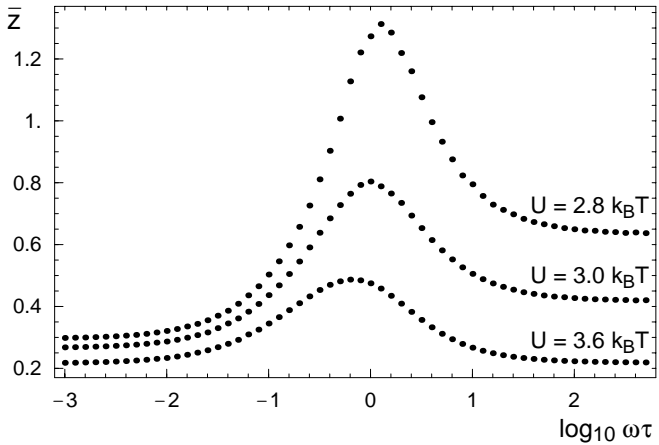}}
\caption{The averaged rescaled separation $\bar{z}$ of the membranes as a function of the mean switching rate $\omega$ for the average fraction $X=0.5$ of `on'-stickers. The membrane separation has a maximum at intermediate switching rates $\omega$, and descreases with increasing sticker binding energy $U$. The data points are obtained from Monte Carlo simulations with up to $10^7$ steps per lattice site on lattices with up to $140\times 140$ sites. Statistical errors are smaller than the symbol sizes. 
\label{resonance}}
\vspace*{-0.3cm}
\end{center} 
\end{figure}
%

{\em Results} --  
The attractive membrane interaction mediated by the `on'-stickers acts against a steric repulsion of the membranes caused by shape fluctuations. If the sticker-mediated attraction is weak, the membrane fluctuations will lead to large membrane separations, and eventually to an unbinding of the membranes. The average membrane separation in the steady state and the location of the unbinding transition therefore reflect the adhesiveness of the membranes. Clearly, the adhesiveness should increase with the sticker binding energy $U$ and the average fraction of  `on'-stickers $X= \omega_{+}/(\omega_{+} + \omega_{-})$. However, even for a constant binding energy and a constant fraction of `on'-stickers, the average membrane separation depends sensitively on the switching rates. In Fig.~\ref{resonance}, the  average membrane separation in the steady state is shown as a function of the mean switching rate $\omega= \frac{1}{2}(\omega_{+} + \omega_{-})$. The fraction of `on'-sticker is $X=0.5$, which implies $\omega_{+}=\omega_{-}$. The average membrane separation exhibits a resonance: it is maximal at intermediate switching rates $\omega$, and considerably smaller at high or low switching rates. This resonance of the membrane separation can be understood from an associated resonance in the escape rate of a single bound sticker molecule, see Appendix.

\begin{figure}[b]
\begin{center}
\resizebox{\columnwidth}{!}{\includegraphics{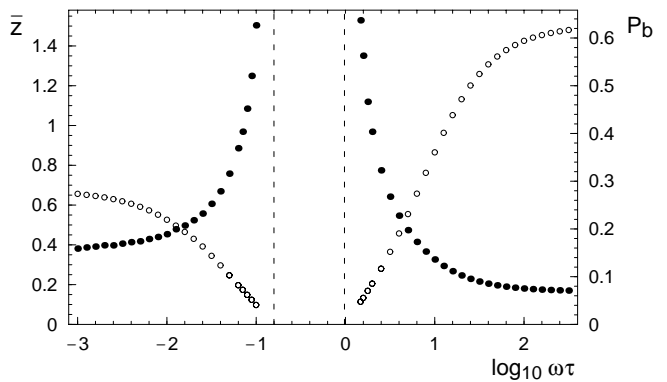}}
\caption{Average membrane separation $\bar{z}$ (filled circles) and contact probability $P_b$ (open circles) as a function of the mean switching rate $\omega$. The fraction of `on'-stickers here is $X=0.2$, and the sticker binding energy is $U=10k_BT$. With increasing switching rate, the membranes first unbind and then rebind at the critical rates $\log \omega_c\tau =-0.79\pm 0.07$ and $\log \omega_c\tau =-0.02\pm 0.05$ (dashed lines), which are obtained from extrapolation to $P_b = 0$ and $1/\bar{z}=0$. \label{reentrant}}
\end{center} 
\vspace*{-0.3cm}
\end{figure}

The resonance of the membrane separation can lead to a reentrant unbinding transition, see Fig.~\ref{reentrant}. For specific values of the fraction $X$ and binding energy $U$ of `on'-stickers, the membranes then first unbind with increasing switching rate $\omega$, and later rebind. The unbinding transitions of the active membranes are continuous. At an unbinding point, the average separation $\bar{z}$ continuously tends to infinity \cite{lipowsky86}, and the contact probability $P_b$, the average fraction of lattice sites with $z_i< z_r$, continuously tends to zero. 

\begin{figure}
\begin{center}
\resizebox{\columnwidth}{!}{\includegraphics{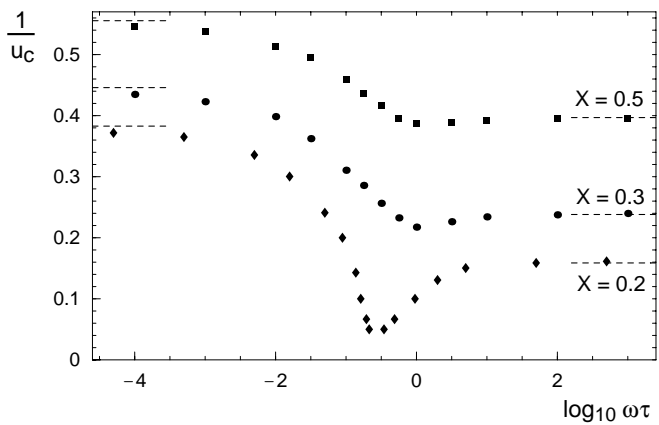}}
\caption{Unbinding lines obtained from Monte Carlo simulations for several values of the fraction $X$ of `on'-stickers. Here, $u_c\equiv U_c/k_B T$ is the dimensionless critical binding energy of the stickers, and $\omega$ is the mean switching rate. At a given value of $X$, the membranes are bound for sticker binding energies $U$ larger than $U_c$ and, thus, for $1/U< 1/U_c$, and unbound for $1/U>1/U_c$. The resonance at intermediate switching rates is is reflected in small values of $1/U_c$ and most pronounced at $X=0.2$. The dashed lines indicate the asymptotic limits given by the eqs.~(\ref{smallomega}) and (\ref{largeomega}). 
\label{unbinding}}
\vspace*{-0.3cm}
\end{center} 
\end{figure}

The steady-state phase behavior of the membranes can be characterized by unbinding lines, see Fig.~{\ref{unbinding}}. The resonance at intermediate switching rates is reflected in minima of the inverse critical binding energy $1/U_c$. This resonance effect increases with decreasing fraction $X$ of `on'-stickers. In the asymptotic limits of small and large switching rates, the unbinding behavior of the active membranes can be related to the behavior of `equilibrium membranes'. In the limit of small switching rates, the active membranes are equivalent to membranes with a concentration $X$ of stickers with `permanent' interaction potential $V(z)$. The effective adhesion potential $V_{\text{ef}}$ of such membranes can be determined exactly by integrating over the stickers degrees of freedom in the partition function \cite{lipowsky96weikl01}. Close to the unbinding point, this effective potential is given by  $\exp\left[-V_{\rm{ef}}(z_i)/k_BT\right] = X \exp\left[-V(z_i)/k_BT\right] +1 -X$. Hence, the critical sticker binding energy is
\begin{equation} 
U_c(\omega = 0)= -k_B T \ln\frac{X}{\exp\left[U_c^{\text{hm}}/k_B T\right] +  X - 1} \label{smallomega}
\end{equation}
where $U_c^{\text{hm}}$ is the critical binding energy of a homogeneous membrane in a square-well potential with the same rescaled potential range $z_r=0.1$. From Monte Carlo simulations, we obtain $U_c^{\text{hm}}=1.26\pm 0.01$. In the limit of fast sticker switching, the critical binding energy behaves as
\begin{equation}
U_c(\omega = \infty) = U_c^{\text{hm}}/X  \label{largeomega}
\end{equation}
since the membranes then move in the average potential $X V(z_i)$.

%

{\em Discussion and conclusions} --
We have shown that active switching processes of stickers can have a strong impact on the separation and unbinding behavior of membranes. The adhesiveness of an `equilibrium sticker' is characterized by its binding energy $U$. In contrast, the adhesiveness of an active sticker depends also sensitively on the transition rates of the conformational switching process.

At low switching rates, a bound sticker can unbind either by (i) `hopping' out of the potential well, or by (ii) diffusing out of the potential range after being switched `off', since the probability for being switched `on' in the subsequent time steps is small. The typical time scales for these processes are (i) the intrinsic unbinding time $t_{\text{ub}}\sim \tau \exp[U]$, and (ii) the inverse `off'-rate $1/\omega_{-}$. For $\omega_{-}\ll 1/t_{\text{ub}}$, the adhesion behavior is nearly unaffected by the sticker switching. The impact of the active switching becomes clearly noticable for switching rates with $\omega_{-}\simeq 1/t_{\text{ub}}$. For biological adhesion molecules such as integrins or selectins, the typical dissociation times $t_{\text{ub}}$ are of the order of seconds \cite{zhang02evans01}.  Switching processes on second or subsecond timescales  therefore should allow to control the adhesion behavior of biological or biomimetic membranes.

For a given concentration $X$ of `on'-stickers, the adhesion strength in our model is minimal at resonant mean switching rates $\omega \sim 1/\tau$, see Figs.~\ref{resonance} and \ref{unbinding}. The characteristic timescale $\tau$ corresponds to the typical relaxation time of membrane segments with linear size $a$, the distance between the stickers. If hydrodynamic damping dominates, this relaxation time scales as $4\eta a^3/(\pi^3 \kappa)$ \cite{brochard75} where $\eta$ is the viscosity of the surrounding solvent. For the typical bending rigidity $\kappa = 10 k_BT$, an average lateral sticker distance of 180~nm, and the typical value $\eta = 0.06$ poise for the cytoplam viscosity of a cell, we obtain a relaxation time scale of 0.1~ms. On this length scale, intermonolayer friction leads to the same estimate for the relaxation time \cite{seifert93}.\\

\begin{acknowledgments}
B.R. thanks Marek Napi\'{o}rkowski for stimulating
discussions.
\end{acknowledgments}

\clearpage

\begin{appendix}
\section{Appendix}

In our simulations, the interaction energy  for a Monte Carlo step at time $t$ and site $i$ with rescaled separation $z_i$ is  $(V(z_i)/\tau) \int_{t-\tau}^t n(t') {\text d}t'$. This interaction energy includes an average over the sticker switching process $n(t')$ in the time interval from $t-\tau$ to $t$. To understand the resonance in the membrane separation observed in our simulations, it is instructive to consider the escape of a single sticker, or single particle, from the dimensionless stochastic potential $v = (u/\tau)\int_0^t n(t') \text{d}t'$ with binding energy $u=U/k_B T$ and $0<v<\infty$. At $t=0$, the sticker is `on'. To determine the average escape rate $k\equiv\langle e^{-v}\rangle$ at time $t=\tau$, we consider the  stochastic differential equation $\text{d}v/\text{d} t = (u/\tau) n(t)$ and construct the corresponding Fokker-Planck equations [17]
\begin{eqnarray} 
\frac{\partial P_0}{\partial t} = \omega_{-} P_1 - \omega_{+}P_0 \hspace*{0.8cm}\\
\frac{\partial P_1}{\partial t} = \omega_{+} P_0 - \omega_{-}P_1 -\frac{u}{\tau}\frac{\partial P_1}{\partial v}
\end{eqnarray}
Here, $P_1(v,t)$ and $P_0(v,t)$ are the probability distributions for $n(t)=1$ and $n(t)=0$, respectively. Multiplying each of these equations with $e^{-v}$ and integrating over $v$ leads to the ordinary first-order differential equations
\begin{equation}
\dot{b}_1 = \omega_{+} b_0 - (\omega_{-} + u/\tau)b_1\, , \;\; \dot{b}_0 = \omega_{-} b_1 - \omega_{+} b_0
\end{equation}
for the two auxiliary variables $b_1(t)=\int e^{-v} P_1(v,t)\text{d}v$ and  $b_0(t)=\int e^{-v} P_0(v,t)\text{d}v$. In terms of these auxiliary variables, the average escape rate is $k=\langle e^{-v}\rangle = b_1(\tau) + b_0(\tau)$. For the initial condition $n(0) = 1$, and thus $b_1(0) =1 $ and $b_0(0) = 0$, we obtain 
\begin{equation}
k= \left(\cosh \Lambda - \frac{u - 2\omega\tau}{2 \Lambda}\sinh \Lambda\right) \exp\left[-\frac{u+2\omega\tau}{2} \right] \label{escaperate}
\end{equation}
with $\Lambda = \frac{1}{2}\sqrt{(u+ 2\omega \tau)^2 - 8 u\omega\tau X}$.

For the fraction $X=0.5$, the escape rate $k$ as a function of $\omega$ is shown in Fig.~6. In the asymptotic limit of slow switching or small $\omega$, the sticker has a high probability to be still in the initial `on'-state. The average potential of the sticker then is identical with its binding energy $u$, and the escape rate simply is $e^{-u}$. In the limit of fast switching or large $\omega$, the sticker will switch very often between the states. The average potential then is $X u=u/2$, and the escape rate $e^{-Xu}=e^{-u/2}$. But at the intermediate switching rates, the average potential depends on the switching process and can vary from values close to zero (if the sticker is switched `off' early in the interval $\tau$ and remains `off') to the value $u$ (if the sticker remains in the `on'-state). The resonance effect at the intermediate switching rates can be understood from the contribution of switching processes with small average potential $v$ to the escape rate $k=\langle e^{-v}\rangle$. The effect is closely related to the `resonant activation' of particles in potentials with a fluctuating barrier [18].\\[0.1cm]

\resizebox{\columnwidth}{!}{\includegraphics{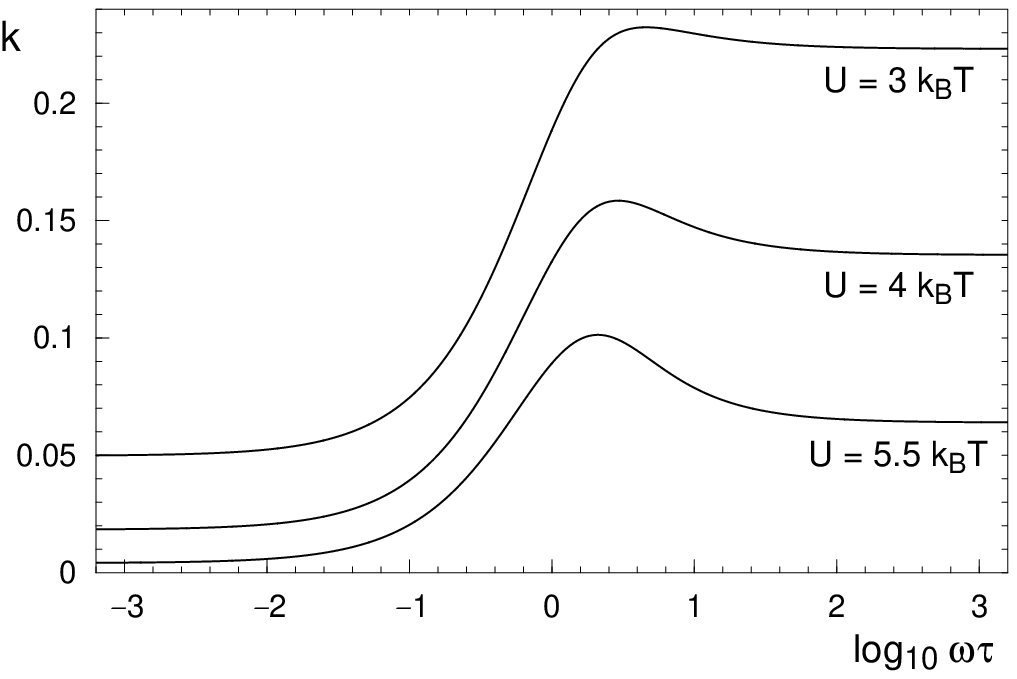}}

\vspace*{-1.4cm}

\begin{figure}[b]
\begin{center}
\caption{Escape rate $k$ for a single sticker as a function of the mean switching rate $\omega$ at the average fraction $X=0.5$ of the `on'-state and several values of the sticker binding energy $U$, as given by eq.~(\ref{escaperate}).
\label{escaperate}}
\end{center} 
\end{figure}

\end{appendix}

\begin{thebibliography}{99}
\bibitem{footnote} More generally, a violation of detailed balance is obtained for $\omega_{-}^{b}= \exp(-c u)\omega_{-}^{u}$  with $c\ne 1$. We consider here the special case $c=0$, but expect very similar results for any $|c|\ll 1$.
\end{thebibliography}
\end{document}